\documentclass[twocolumn,showpacs,showkeys,preprintnumbers,amsmath,amssymb]{revtex4}

\usepackage{graphicx}

\begin{document}


\title{Study on the Solutions of the Sunyaev-Zeldovich Effect for Clusters of Galaxies}
\author{Satoshi Nozawa}
 \email{snozawa@josai.ac.jp}
\affiliation{
Josai Junior College, 1-1 Keyakidai, Sakado-shi, Saitama, 350-0295,
Japan}

\author{Yasuharu Kohyama and Naoki Itoh}
\affiliation{
Department of Physics, Sophia University, 7-1 Kioi-cho, Chiyoda-ku,
Tokyo, 102-8554, Japan}

\date{\today}

\begin{abstract}
  Based upon the rate equations for the photon distribution function obtained in the previous paper, we study the formal solutions in three different representation forms for the Sunyaev-Zeldovich effect.  By expanding the formal solution in the operator representation in powers of both the derivative operator and electron velocity, we derive a formal solution that is equivalent to the Fokker-Planck expansion approximation.  We extend the present formalism to the kinematical Sunyaev-Zeldovich effect.  The properties of the frequency redistribution functions are studied.  We find that the kinematical Sunyaev-Zeldovich effect is described by the redistribution function related to the electron pressure.  We also solve the rate equations numerically.  We obtain the exact numerical solutions, which include the full-order terms in powers of the optical depth.
\end{abstract}



\maketitle

\section{Introduction}

  The Sunyaev-Zeldovich (SZ) effect\cite{suny72,suny80a,suny80b,suny81}, which arises from the Compton scattering of the cosmic microwave background (CMB) photons by hot electrons in clusters of galaxies (CG), provides a useful method for studies of cosmology.  For the reviews, for example, see Birkinshaw\cite{birk99} and Carlstrom, Holder, and Reese\cite{carl02}.  The original SZ formula has been derived from the Kompaneets equation\cite{komp57} in the nonrelativistic approximation.  However, recent x-ray observations (for example, Schmidt et al.\cite{tuck98} and Allen et al.\cite{alle02}) have revealed the existence of high-temperature CG such as $k_{B} T_{e} \simeq $20keV.  Wright\cite{wrig79} and Rephaeli and his collaborator\cite{reph95, reph97} have done pioneering work including the relativistic corrections to the SZ effect for the CG.

In the last ten years remarkable progress has been made in theoretical studies of the relativistic corrections to the SZ effects for the CG.  Stebbins\cite{steb97} generalized the Kompaneets equation.  Challinor and Lasenby\cite{chal98} and Itoh, Kohyama, and Nozawa\cite{itoh98} have adopted a relativistically covariant formalism to describe the Compton scattering process and have obtained higher-order relativistic corrections to the thermal SZ effect in the form of the Fokker-Planck approximation.  Nozawa, Itoh, and Kohyama\cite{noza98} have extended their method to the case where the CG is moving with a peculiar velocity with respect to the CMB and have obtained the relativistic corrections to the kinematical SZ effect.  Their results were confirmed by Challinor and Lasenby\cite{chal99} and also by Sazonov and Sunyaev\cite{sazo98a, sazo98b}.  Itoh, Nozawa, and Kohyama\cite{itoh00} have also applied the covariant formalism to the polarization SZ effect\cite{suny80b,suny81}.  Itoh and his collaborators (including the present authors) have done extensive studies on the SZ effects, which include the double scattering effect\cite{itoh01}, the effect of the motion of the observer\cite{noza05}, high precision analytic fitting formulae to the direct numerical integrations\cite{noza00,itoh02} and high precision calculations\cite{itoh04,noza06}.  The importance of the relativistic corrections is also exemplified through the possibility of directly measuring the cluster temperature using purely the SZ effect\cite{hans04}.

  On the other hand, the SZ effect in the CG has been studied also for the nonthermal distributions by several groups\cite{enss00,cola03,cola09}.  The nonthermal distribution functions, for example, the power-law distributions, have a long tail in high electron energy regions.  Therefore the relativistic corrections for the SZ effect could be more important than the thermal distribution.

  The relativistic SZ effect has been studied in several different approaches.  Wright\cite{wrig79} and Rephaeli\cite{reph95} calculated the photon frequency redistribution function in the electron rest frame using the scattering probability derived by Chandrasekhar\cite{chan50}, which is called as Wright's method in the present paper.  Another approach is the relativistic generalization of the Kompaneets equation\cite{komp57}, where the relativistically covariant Boltzmann collisional equation is solved for the photon distribution function.  This approach was used by Challinor and Lasenby\cite{chal98} and Itoh, Kohyama, and Nozawa\cite{itoh98}, which is called the covariant formalism in the present paper.  Although the two are very different approaches, the obtained results for the SZ effect agreed extremely well.  This has been a longstanding puzzle in the field of the relativistic study of the SZ effect for the last ten years.

 Very recently, Shimon and Rephaeli\cite{shim04}, Boehm and J. Lavalle\cite{boeh08} and Nozawa and Kohyama\cite{noza09} (denoted NK hereafter) discussed the equivalence between different approaches.  In particular, NK showed that Wright's method and the covariant formalism were indeed mathematically equivalent in the approximation of the Thomson limit, which is fully valid for the CMB photon energies.  This explained the reason why the two different approaches produced same results for the SZ effect even in the relativistic energies for electrons.  Thus, NK clarified the situation of the longstanding puzzle.

 The third method for the study of the SZ effect is the direct numerical integration of the rate equation of the photon spectral distortion function.  The first-order calculation in powers of the optical depth $\tau$ was done by Itoh, Kohyama, and Nozawa\cite{itoh98} for $\tau \ll 1$.  The exact calculation which includes the full-order terms in powers of $\tau$ was done by Dolgov et al.\cite{dolg01} for $\tau \gg 1$.  The rate equation in the present formalism is expressed in the form that is suitable to the direct numerical calculation.  In the present paper, we solve the rate equations numerically, and show the exact numerical solutions.

  In the present paper, we explore the formal solutions for the rate equations in the following representation forms: the multiple scattering representation\cite{noza09}, operator representation\cite{bern90}, and Fourier transform representation\cite{tayl89}.  In particular, we show that these representation forms are identical.  With the operator representation form, we derive a formal solution that is equivalent to the formal solution obtained by the Fokker-Planck expansion approximation.

  The present paper is organized as follows:  Starting from the rate equations derived in the NK paper, we derive in Sec.~II the formal solutions of the rate equations for the SZ effect.  In Sec.~III, the present formalism is extended to the case which includes the peculiar velocity of the CG.  In Sec.~IV, we solve the rate equations numerically, and obtain exact numerical solutions for the thermal SZ effect and kinematical SZ effect.  The numerical solutions for the nonthermal electron distributions are also presented in Sec.~IV.  Finally, concluding remarks are given in Sec.~V.

\section{Formal Solutions for the thermal Sunyaev-Zeldovich Effect}

\subsection{Rate equations}

  In the NK paper, it was shown that the covariant formalism\cite{itoh98} and Wright's method\cite{wrig79} were mathematically equivalent in the approximation $\gamma \omega/m \ll 1$ which is fully valid for the CMB photon energies $\omega$, where $\gamma = 1/\sqrt{1-\beta^{2}}$, $\beta$ and $m$ are the velocity and rest mass of the electron, respectively.  The two formalisms were connected each other by the Lorentz transformations for the zenith angles.  Readers may be referred to the NK paper\cite{noza09} for the details.  In the present paper, we use the expression in Wright's method for the photon frequency redistribution function, which was derived in the NK paper.

 In the present section, we consider the case that both of the CG and observer are fixed to the CMB frame.  As a reference system, we choose the system that is fixed to the CMB.  (Three frames are identical in the present case.)  Throughout this paper, we use the natural unit $\hbar = c = 1$, unless otherwise stated explicitly.

  The rate equations for the photon distribution function $n(x)$ and the spectral intensity function $I(x)$ were derived in the NK paper, where $x=\omega/k_{B}T_{CMB}$ is the photon energy in units of the thermal energy of the CMB.   We recall the results here to make the present paper more self-contained.  They are given as follows\cite{noza09}:
\begin{eqnarray}
&&\hspace{-10mm}
\frac{\partial n(x)}{\partial \tau}
 = \int_{-\infty}^{\infty}ds P_1(s)
\left[n(e^sx)- n(x)\right] \, ,
\label{eq2a-1}   \\
&&\hspace{-10mm}
\frac{\partial I(x)}{\partial \tau}
 = \int_{-\infty}^{\infty}ds
{P}_1(s) \left[I(e^{-s}x)- I(x)\right]  \, ,
\label{eq2a-2}  \\
&&\hspace{+12mm}
d\tau  =  n_e\sigma_T dt \, , 
\label{eq2a-3}
\end{eqnarray}
where $I(x)=I_{0}x^{3}n(x)$, $I_{0}=(k_{B}T_{CMB})^{3}/2\pi^{2}$, $n_{e}$ is the electron number density, $\sigma_{T}$ is the Thomson scattering cross section.  In Eqs.~(\ref{eq2a-1}) and (\ref{eq2a-2}), $P_{1}(s)$ is the redistribution function for photon of a frequency shift $s$, which is defined by $e^{s}=x^{\prime}/x$,
\begin{eqnarray}
&&\hspace{-10mm}
P_1(s) = \int_{\beta_{min}}^{1}d\beta\beta^2\gamma^5 \tilde{p}_e(E)P(s,\beta)  \, ,
\label{eq2a-4}  \\
&&\hspace{-10mm}
P(s,\beta) = \frac{e^{s}}{2\beta\gamma^4}
\int_{\mu_1(s)}^{\mu_2(s)}d\mu_0 \frac{1}{(1-\beta\mu_0)^2} f\left(\mu_0, \mu_0^{\prime} \right)   \, ,
\label{eq2a-5}
\end{eqnarray}
\begin{eqnarray}
&&\hspace{-10mm}
f(\mu_0,\mu_0^{\prime}) = \frac{3}{8}\left[
  1 + \mu_0^2\mu_0^{\prime 2}+\frac{1}{2}(1-\mu_0^2)(1-\mu_0^{\prime 2})
\right]  \, ,
\label{eq2a-6}
\end{eqnarray}
where $\tilde{p}_{e}(E) \equiv m^{3}p_{e}(E)$ is the electron distribution function of a momentum $p$ which is normalized by $\int_{0}^{\infty} dp p^{2} p_{e}(E)=1$.  The total probability is
\begin{eqnarray}
&&\hspace{-10mm}
\int_{-\infty}^{\infty}ds P_1(s) = 1 \, .
\label{eq2a-7}
\end{eqnarray}
Variables appearing in Eqs.~(\ref{eq2a-4}) -- (\ref{eq2a-6}) are summarized as follows:
\begin{eqnarray}
&&\hspace{-10mm}
\beta_{min} = (1-e^{-|s|})/(1+e^{-|s|})  \, ,
\label{eq2a-8} \\
&&\hspace{-10mm}
\mu_{0}^{\prime} = [1-e^s(1-\beta\mu_0)]/\beta  \, ,
\label{eq2a-9}  \\
&&\hspace{-10mm}
\mu_1(s) = \left\{
\begin{array}{ll}
-1 &\quad  {\rm for} \, \, \, s \leq 0 \\
{[1-e^{-s}(1+\beta)]/\beta} &\quad {\rm for} \, \, \, s > 0
\end{array}
\right.  \, ,
\label{eq2a-10} \\
&&\hspace{-10mm}
\mu_2(s) = \left\{
\begin{array}{ll}
{[1-e^{-s}(1-\beta)]/\beta} &\quad {\rm for} \, \, \, s < 0 \\
1 &\quad  {\rm for} \, \, \, s \geq 0 
\end{array}
\right. \, .
\label{eq2a-11}
\end{eqnarray}
It should be noted that the following useful relations
\begin{eqnarray}
&&\hspace{-10mm}
P(s, \beta)e^{-3s} = P(-s,\beta) \, , \, \, \,P_{1}(s)e^{-3s} = P_{1}(-s) \,
\label{eq2a-12}
\end{eqnarray}
were used in deriving Eq.~(\ref{eq2a-2}).

\subsection{Formal solutions}

  In the present section, we explore the formal solutions of the rate equations of Eqs.~(\ref{eq2a-1}) and (\ref{eq2a-2}) in the following representation forms:{\it the multiple scattering representation}, {\it operator representation}, and {\it Fourier transform representation}.  In particular, we show that these representation forms are identical.

  First, it is familiar to express the formal solutions in the multiple scattering representation.  They are given, for example, in the NK paper\cite{noza09} as follows:
\begin{eqnarray}
&&\hspace{-10mm}
n(x) = \int_{-\infty}^{\infty}ds
P(s) n_{0}(e^{s}x)  \, ,
\label{eq2b-1}  \\
&&\hspace{-10mm}
I(x) = \int_{-\infty}^{\infty}ds
P(s) I_{0}(e^{-s}x)  \, ,
\label{eq2b-2}
\end{eqnarray}
where $n_{0}(x)$ and $I_{0}(x)$ are the initial functions at $\tau$=0.  In Eqs.~(\ref{eq2b-1}) and (\ref{eq2b-2}), $P(s)$ is expressed by
\begin{eqnarray}
&&\hspace{-10mm}
P(s) = e^{-\tau} \sum_{j=0}^{\infty} \frac{\tau^{j}}{j!} P_{j}(s) \, ,
\label{eq2b-3}  \\
&&\hspace{-10mm}
P_{0}(s) = \delta(s)  \, ,
\label{eq2b-4}  \\
&&\hspace{-10mm}
P_{j}(s) = \int_{-\infty}^{\infty}ds_1 P_1(s_1)\cdots
\int_{-\infty}^{\infty}ds_{j-1} P_1(s_{j-1}) 
\nonumber  \\
&&\hspace{+30mm}
\times P_1\left(s-\sum_{i=1}^{j-1}s_i \right) \, ,
\label{eq2b-5}
\end{eqnarray}
where $P_{j}(s)$ is the redistribution function for the multiple scattering of the $j$-th order.

  Next, let us derive the formal solutions in the operator representation.  This method has an advantage that not only the solution can be expressed in a concise form but also it is useful to study relations between different methods.  We write the following useful operator identity introduced by Bernstein and Dodelson\cite{bern90}:
\begin{eqnarray}
&&\hspace{-10mm}
e^{\lambda D} f(x) e^{-\lambda D} = f(e^{\lambda}x)  \, ,
\label{eq2b-6}  \\
&&\hspace{+7.5mm}
D \equiv x \frac{\partial}{\partial x}  \, ,
\label{eq2b-7}
\end{eqnarray}
where $\lambda$ is a number and $f(x)$ is an arbitrary function of $x$.  Note that the $e^{-\lambda D}$ factor goes to 1 since the derivative has nothing to act on.  It is important to understand the physical meaning of the operator $D$.  Inserting $\lambda=s$ and $f(x)=x$ into Eq.~(\ref{eq2b-6}), one finds
\begin{eqnarray}
&&\hspace{-10mm}
e^{s D} x e^{-s D} = e^{s}x = x^{\prime}  \, ,
\label{eq2b-7.5}
\end{eqnarray}
where the definition of $s$ ($e^{s}=x^{\prime}/x$) was used.  Therefore, the operator $D$ is nothing but the shift operator for $x$ in the present (Wright's) formalism.

Introducing an operator $\mathcal{O}(\pm D)$ by
\begin{eqnarray}
&&\hspace{-10mm}
\mathcal{O}(\pm D) \equiv \int_{-\infty}^{\infty}ds P_1(s) e^{\pm sD} -1 \, ,
\label{eq2b-8}
\end{eqnarray}
and inserting into Eqs.~(\ref{eq2a-1}) and (\ref{eq2a-2}), one obtains
\begin{eqnarray}
&&\hspace{-10mm}
\frac{\partial n(x)}{\partial \tau}
 = \mathcal{O}(D) n(x)  \, ,
\label{eq2b-9}  \\
&&\hspace{-10mm}
\frac{\partial I(x)}{\partial \tau}
 = \mathcal{O}(-D) I(x)  \, ,
\label{eq2b-10}
\end{eqnarray}
where Eqs.~(\ref{eq2a-7}) and (\ref{eq2b-6}) were used in the derivation.  Thus, the following formal solutions in the operator representation are obtained:
\begin{eqnarray}
&&\hspace{-10mm}
n(x) = e^{\tau \mathcal{O}(D)} n_{0}(x) \, ,
\label{eq2b-11}  \\
&&\hspace{-10mm}
I(x) = e^{\tau \mathcal{O}(-D)} I_{0}(x) \, .
\label{eq2b-12}
\end{eqnarray}
It should be noted that Eqs.~(\ref{eq2b-9}) and (\ref{eq2b-10}) require the following condition for $\mathcal{O}(D)$:
\begin{eqnarray}
&&\hspace{-10mm}
x^{3} \mathcal{O}(D) =  \mathcal{O}(-D) x^{3}  \, .
\label{eq2b-13add}
\end{eqnarray}
In the present case, Eq.~(\ref{eq2b-13add}) is guaranteed to be valid by Eq.~(\ref{eq2a-12}).

  We now show that Eqs.~(\ref{eq2b-11}) and (\ref{eq2b-12}) are identical to Eqs.~(\ref{eq2b-1}) and (\ref{eq2b-2}), respectively.  Let us start with Eq.~(\ref{eq2b-11}) as follows:
\begin{eqnarray}
&&\hspace{-10mm}
n(x) = {\rm exp} \left\{ \tau \left[ \int_{-\infty}^{\infty}ds P_1(s) e^{sD} - 1 \right] \right\} n_{0}(x)  \, ,
\nonumber  \\
&&\hspace{-1mm}
=e^{-\tau} \sum_{j=0}^{\infty} \frac{\tau^{j}}{j!} \left( \int_{-\infty}^{\infty}ds P_1(s) e^{sD} \right)^{j} n_{0}(x)  \, .
\label{eq2b-13}
\end{eqnarray}
Applying the identity of Eq.~(\ref{eq2b-6}), the $j$-th order term ($j \geq 1$) can be rewritten as
\begin{eqnarray}
&&\hspace{-10mm}
\left( \int_{-\infty}^{\infty}ds P_1(s) e^{sD} \right)^{j} n_{0}(x)
\nonumber  \\
&&\hspace{-10mm}
= \int_{-\infty}^{\infty}ds_1 P_1(s_1)\cdots
\int_{-\infty}^{\infty}ds_{j} P_1(s_{j}) n_{0}(e^{s_{1} +\cdots + s_{j}}x) \, ,
\nonumber  \\
&&\hspace{-10mm}
= \int_{-\infty}^{\infty}ds_1 P_1(s_1)\cdots
\int_{-\infty}^{\infty}ds P_1 \left(s - \sum_{i=1}^{j-1} s_{i} \right) n_{0}(e^{s}x)  \, ,
\nonumber  \\
&&\hspace{-10mm}
= \int_{-\infty}^{\infty}ds P_{j}(s) n_{0}(e^{s}x)  \, .
\label{eq2b-14}
\end{eqnarray}
Inserting Eq.~(\ref{eq2b-14}) into Eq.~(\ref{eq2b-13}), one finally obtains Eq.~(\ref{eq2b-1}).  Thus, it has been shown that Eq.~(\ref{eq2b-11}) is identical to Eq.~(\ref{eq2b-1}).  The equivalence between Eqs.~(\ref{eq2b-2}) and (\ref{eq2b-12}) can be also shown in a similar manner.

  Let us now derive the third formal solutions for the rate equations.  The Fourier transform representation was introduced by Taylor and Wright\cite{tayl89}, where the redistribution function was derived in their Eq.~(28).  Consider the Fourier transform for $P(s)$ of Eq.~(\ref{eq2b-3}) as follows:
\begin{eqnarray}
&&\hspace{-10mm}
\bar{P}(k) \equiv \int_{-\infty}^{\infty}ds P(s) e^{-iks}  \, ,
\nonumber  \\
&&\hspace{-1mm}
=e^{-\tau} \sum_{j=0}^{\infty} \frac{\tau^{j}}{j!} \int_{-\infty}^{\infty}ds P_{j}(s) e^{-iks}  \, .
\label{eq2b-15}
\end{eqnarray}
Using Eq.~(\ref{eq2b-5}), the $j$-th order term in Eq.~(\ref{eq2b-15}) is written by
\begin{eqnarray}
&&\hspace{-10mm}
\int_{-\infty}^{\infty}ds P_{j}(s) e^{-iks} = \int_{-\infty}^{\infty}ds_1 P_1(s_1)\cdots \int_{-\infty}^{\infty}ds_{j} P_1(s_{j}) 
\nonumber  \\
&&\hspace{+21mm}
\times e^{-ik(s_{1} +\cdots + s_{j})} \, ,
\nonumber  \\
&&\hspace{+18mm}
= \left( \int_{-\infty}^{\infty}ds P_{1}(s) e^{-iks} \right)^{j}  \, .
\label{eq2b-16}
\end{eqnarray}
Inserting Eq.~(\ref{eq2b-16}) into Eq.~(\ref{eq2b-15}), one finally obtains the Fourier transform,
\begin{eqnarray}
&&\hspace{-10mm}
\bar{P}(k) = e^{\tau [ \bar{P}_{1}(k) - 1 ]}  \, ,
\label{eq2b-17}  \\
&&\hspace{-10mm}
\bar{P}_{1}(k) \equiv \int_{-\infty}^{\infty}ds P_{1}(s) e^{-iks}  \, .
\label{eq2b-18}
\end{eqnarray}
The inverse Fourier transform for $\bar{P}(k)$ is
\begin{eqnarray}
&&\hspace{-10mm}
P(s) = \frac{1}{2 \pi} \int_{-\infty}^{\infty}dk e^{\tau [ \bar{P}_{1}(k) - 1 ]} e^{iks}  \, .
\label{eq2b-19}
\end{eqnarray}
Inserting Eq.~(\ref{eq2b-19}) into Eq.~(\ref{eq2b-1}), one obtains
\begin{eqnarray}
&&\hspace{-10mm}
n(x) = \frac{1}{2 \pi} 
 \int_{-\infty}^{\infty}dk e^{\tau [ \bar{P}_{1}(k) - 1 ]} \int_{-\infty}^{\infty}ds e^{iks} n_{0}(e^{s}x)  \, ,
\nonumber  \\
&&\hspace{-2mm}
=\int_{-\infty}^{\infty}dk e^{\tau [ \bar{P}_{1}(k) - 1 ]} \delta(k-iD) n_{0}(x) \, ,
\nonumber  \\
&&\hspace{-2mm}
= e^{\tau [ \bar{P}_{1}(iD) - 1 ]} n_{0}(x)  \, ,
\label{eq2b-20}
\end{eqnarray}
where the operator identity of Eq.~(\ref{eq2b-6}) was used in the derivation.  Thus, the formal solution for $n(x)$ in the Fourier transform representation has been obtained.  Similarly, the formal solution for $I(x)$ is also obtained as follows:
\begin{eqnarray}
&&\hspace{-10mm}
I(x) = e^{\tau [ \bar{P}_{1}(-iD) - 1 ]} I_{0}(x)  \, .
\label{eq2b-21}
\end{eqnarray}

  Here, we show that Eqs.(\ref{eq2b-20}) and (\ref{eq2b-21}) are identical to Eqs.~(\ref{eq2b-11}) and (\ref{eq2b-12}), respectively.  In deriving Eq.~(\ref{eq2b-20}), we found that $\delta(k-iD)$ was valid.  By putting
\begin{eqnarray}
&&\hspace{-10mm}
k_{D} \equiv i D  \, ,
\label{eq2b-22}
\end{eqnarray}
and inserting into Eq.~(\ref{eq2b-8}), one obtains
\begin{eqnarray}
&&\hspace{-10mm}
\mathcal{O}(\pm D) = \int_{-\infty}^{\infty}ds P_1(s) e^{\mp ik_{D}s} -1 \, ,
\nonumber  \\
&&\hspace{+2mm}
= \bar{P}_{1}(\pm k_{D}) - 1  \, ,
\nonumber  \\
&&\hspace{+2mm}
= \bar{P}_{1}(\pm iD) - 1  \, ,
\label{eq2b-23}
\end{eqnarray}
where Eq.~(\ref{eq2b-18}) was used in the derivation.  Inserting Eq.~(\ref{eq2b-23}) into Eqs.~(\ref{eq2b-11}) and (\ref{eq2b-12}), one finally obtains Eqs.(\ref{eq2b-20}) and (\ref{eq2b-21}).  Thus, the equivalence between the operator representation and Fourier transform representation has been shown.  Hence, the three representation forms are identical.

\subsection{Fokker-Planck expansion approximation}

  Using the operator representation method, we explore the relation of the formal solutions between the Fokker-Planck expansion approximation and the present method.  For simplicity, we restrict ourselves to the SZ effect in the present paper, however, the extension to the kinematical SZ effect is straightforward.

  In Itoh, Kohyama, and Nozawa\cite{itoh98} (denoted by IKN hereafter), the rate equation for the photon distribution function was expressed by their Eqs.~(2.10)--(2.21) in the Fokker-Planck expansion approximation.  Keeping the leading-order terms, the rate equation is reexpressed as follows:
\begin{eqnarray}
&&\hspace{-10mm}
\frac{ \partial n(x)}{ \partial \tau} = \left[
   x \frac{ \partial}{ \partial x}  I_{1} 
 + x^{2} \frac{ \partial^{2}}{ \partial x^{2}} I_{2}
 + x^{3} \frac{ \partial^{3}}{ \partial x^{3}} I_{3}
 + x^{4} \frac{ \partial^{4}}{ \partial x^{4}} I_{4}
 \right.  \nonumber \\
 &&\hspace{+5mm}
 + x^{5} \frac{ \partial^{5}}{ \partial x^{5}} I_{5}
 + x^{6} \frac{ \partial^{6}}{ \partial x^{6}} I_{6}
 + x^{7} \frac{ \partial^{7}}{ \partial x^{7}} I_{7}
 + x^{8} \frac{ \partial^{8}}{ \partial x^{8}} I_{8}
\nonumber \\
&&\hspace{+5mm}
\left.
 + x^{9} \frac{ \partial^{9}}{ \partial x^{9}} I_{9}
 + x^{10}\frac{ \partial^{10}}{ \partial x^{10}} I_{10} \right] n(x)  \, ,
\label{eq2c-1}
\end{eqnarray}
where $I_{1}, I_{2}, \cdots I_{10}$ are defined by Eqs.~(2.12)--(2.21) of the IKN paper.  The explicit forms are
\begin{eqnarray}
&&\hspace{-10mm}
I_{1} = 4 \theta_{e} + 10 \theta_{e}^{2} + \frac{15}{2} \theta_{e}^{3}
  - \frac{15}{2} \theta_{e}^{4} + \frac{135}{32} \theta_{e}^{5} \, ,
\label{eq2c-2}  \\
&&\hspace{-10mm}
I_{2} = \theta_{e} + \frac{47}{2} \theta_{e}^{2} + \frac{1023}{8} \theta_{e}^{3} + \frac{2505}{8} \theta_{e}^{4} + \frac{30375}{128} \theta_{e}^{5}  \, ,
\label{eq2c-3}  \\
&&\hspace{-10mm}
I_{3} =  \frac{42}{5} \theta_{e}^{2} + \frac{868}{5} \theta_{e}^{3} + \frac{7098}{5} \theta_{e}^{4} + \frac{62391}{10} \theta_{e}^{5}  \, ,
\label{eq2c-4}  \\
&&\hspace{-10mm}
I_{4} = \frac{7}{10} \theta_{e}^{2} + \frac{329}{5} \theta_{e}^{3} + \frac{14253}{10} \theta_{e}^{4} + \frac{614727}{40} \theta_{e}^{5}  \, ,
\label{eq2c-5}  \\
&&\hspace{-10mm}
I_{5} = \frac{44}{5} \theta_{e}^{3} + \frac{18594}{35} \theta_{e}^{4} + \frac{124389}{10} \theta_{e}^{5}  \, ,
\label{eq2c-6}  \\
&&\hspace{-10mm}
I_{6} = \frac{11}{30} \theta_{e}^{3}  + \frac{12059}{140} \theta_{e}^{4} + \frac{355703}{80} \theta_{e}^{5}  \, ,
\label{eq2c-7}  \\
&&\hspace{-10mm}
I_{7} = \frac{128}{21} \theta_{e}^{4} + \frac{16568}{21} \theta_{e}^{5} \, ,
\label{eq2c-8}  \\
&&\hspace{-10mm}
I_{8} = \frac{16}{105} \theta_{e}^{4} + \frac{7516}{105} \theta_{e}^{5} \, ,
\label{eq2c-9}  \\
&&\hspace{-10mm}
I_{9} = \frac{22}{7} \theta_{e}^{5}  \, ,
\label{eq2c-10}  \\
&&\hspace{-10mm}
I_{10} = \frac{11}{210} \theta_{e}^{5}  \, ,
\label{eq2c-11}
\end{eqnarray}
where $x=\omega/k_{B}T_{CMB}$ and $\theta_{e}=k_{B}T_{e}/mc^{2}$.  It should be noted as follows:  In Eqs.~(2.12)--(2.21) of the IKN paper, terms not appearing in Eqs.~(\ref{eq2c-2})--(\ref{eq2c-11}) are $O(T_{CMB}/T_{e}) \ll 1$, therefore,  they are safely neglected.

  We now rewrite the rate equation of Eq.~(\ref{eq2c-1}) in terms of the operator $D$ defined by Eq.~(\ref{eq2b-7}).  One obtains as follows:
\begin{eqnarray}
&&\hspace{-10mm}
\frac{\partial n(x)}{\partial \tau}
 = \mathcal{O}_{IKN}(D) n(x)   \, ,
\label{eq2c-12}  \\
&&\hspace{-10mm}
\mathcal{O}_{IKN}(D) = \sum_{n=1}^{5} \mathcal{O}_{n} \theta_{e}^{n} \, ,
\label{eq2c-13}
\end{eqnarray}
where
\begin{eqnarray}
&&\hspace{-50mm}
\mathcal{O}_{1} = 3 D + D^{2}  \, ,
\label{eq2c-14}
\end{eqnarray}
\begin{eqnarray}
&&\hspace{-15mm}
\mathcal{O}_{2} = -\frac{9}{10} D + 6 D^{2} + \frac{21}{5} D^{3} + \frac{7}{10} D^{4}  \, ,
\label{eq2c-15}
\end{eqnarray}
\begin{eqnarray}
&&\hspace{-10mm}
\mathcal{O}_{3} = -\frac{31}{40} D - \frac{1039}{120} D^{2} + \frac{43}{10} D^{3} + \frac{269}{30} D^{4} 
\nonumber  \\
&&\hspace{+7mm}
+ \frac{33}{10} D^{5} + \frac{11}{30} D^{6}  \, ,
\label{eq2c-16}
\end{eqnarray}
\begin{eqnarray}
&&\hspace{-4mm}
\mathcal{O}_{4} = \frac{69}{56} D + \frac{575}{56} D^{2} - \frac{2959}{140} D^{3} - \frac{5993}{420} D^{4}
\nonumber  \\
&&\hspace{+7mm}
+ \frac{1011}{140} D^{5} + \frac{605}{84} D^{6} + \frac{64}{35} D^{7} + \frac{16}{105} D^{8}  \, ,
\label{eq2c-17}
\end{eqnarray}
\begin{eqnarray}
&&\hspace{-2mm}
\mathcal{O}_{5} = \frac{4371}{896} D - \frac{81707}{4480} D^{2} + \frac{30339}{560} D^{3} + \frac{4551}{560} D^{4}
\nonumber  \\
&&\hspace{+7mm}
- \frac{3273}{80} D^{5} - \frac{3443}{240} D^{6} + \frac{199}{35} D^{7} + \frac{421}{105} D^{8} 
\nonumber  \\
&&\hspace{+7mm}
+ \frac{11}{14} D^{9} + \frac{11}{210} D^{10}  \, .
\label{eq2c-18}
\end{eqnarray}
Thus, the formal solution in the Fokker-Planck expansion approximation is obtained by
\begin{eqnarray}
&&\hspace{-10mm}
n_{IKN}(x) = e^{\tau \mathcal{O}_{IKN}(D)} n_{0}(x)   \, ,
\label{eq2c-19}
\end{eqnarray}
where $n_{0}(x)$ is the initial solution at $\tau=0$.

  On the other hand, the formal solution in the operator representation was given by
\begin{eqnarray}
&&\hspace{-10mm}
n(x) = e^{\tau \mathcal{O}(D)} n_{0}(x)  \, ,
\label{eq2c-20} \\
&&\hspace{-10mm}
\mathcal{O}(D) = \int_{-\infty}^{\infty}ds P_1(s) e^{sD} -1 \, .
\label{eq2c-21}
\end{eqnarray}
By expanding $e^{sD}$ in Eq.~(\ref{eq2c-21}), one obtains the following:
\begin{eqnarray}
&&\hspace{-10mm}
\mathcal{O}(D) = \sum_{n=1}^{\infty} \frac{D^{n}}{n!} \int_{-\infty}^{\infty}ds P_1(s) s^{n}  \, ,
\nonumber  \\
&&\hspace{0mm}
\equiv \sum_{n=1}^{\infty} d_{n} D^{n}   \, ,
\label{eq2c-22}
\end{eqnarray}
where
\begin{eqnarray}
&&\hspace{-10mm}
d_{n} = \frac{1}{n!} \int_{-\infty}^{\infty}ds P_1(s) s^{n}  \, ,
\nonumber \\
&&\hspace{-5mm}
= \frac{1}{n!} \int_{0}^{\infty}d\tilde{p} \tilde{p}^{2} \tilde{p}_{e}(\beta) \int_{-1}^{+1}d\mu_0 \int_{-1}^{+1}d\mu_0^{\prime} 
\nonumber  \\
&&\hspace{0mm}
\times \frac{1}{2\gamma^4} \frac{1}{(1-\beta\mu_0)^3} 
f\left(\mu_0, \mu_{0}^{\prime} \right) 
\left[ {\rm log} \frac{1- \beta \mu_{0}^{\prime}}{1- \beta \mu_{0}} \right]^{n} \, .
\label{eq2c-23}
\end{eqnarray}
In deriving Eq.~(\ref{eq2c-23}), the definition of $P_{1}(s)$ and $e^{s}=(1-\beta \mu_{0}^{\prime})/(1-\beta \mu_{0})$ were used.  Our task is to calculate Eq.~(\ref{eq2c-23}) for $n=1$ to 10, and compare with Eq.~(\ref{eq2c-13}).  The calculation is performed by expanding all functions in Eq.~(\ref{eq2c-23}) in powers of $\beta$ and keeping up to $O(\beta^{10})$ terms.  In the calculation, we used the symbolic manipulation program {\it Mathematica}.  Then, the integration of Eq.~(\ref{eq2c-23}) was also done analytically.  The result is expressed by the power-series of $\theta_{e}$.  Finally, one has
\begin{eqnarray}
&&\hspace{-10mm}
\mathcal{O}(D) = \mathcal{O}_{IKN}(D)  + O(\theta_{e}^{6}) \, .
\label{eq2c-24}
\end{eqnarray}
Thus, we have obtained a formal solution that is equivalent to the solution in the Fokker-Planck expansion approximation.  Similarly, the formal solution for the spectral intensity function in the Fokker-Planck expansion approximation is
\begin{eqnarray}
&&\hspace{-10mm}
I_{IKN}(x) = e^{\tau \mathcal{O}_{IKN}(-D)} I_{0}(x) \, ,
\label{eq2c-25}
\end{eqnarray}
where $I_{0}(x)$ is the initial solution at $\tau$=0.  It should be noted that the operator $\mathcal{O}_{IKN}(D)$ satisfies the condition of Eq.~(\ref{eq2b-13add}).

\section{Formal Solutions for the kinematical Sunyaev-Zeldovich Effect}

\subsection{Rate equations}

  Let us now consider the case that the CG is moving with a peculiar velocity $\vec{\beta}_{c}$ (=$\vec{v}_{c}/c$) with respect to the CMB.  As a reference system, we choose the system that is fixed to the CMB.  The $z$ axis is fixed to a line connecting the observer and the center of mass of the CG.  (We assume that the observer is fixed to the CMB frame.)  In the present paper we choose the positive direction of the $z$ axis as the direction of the propagation of a photon from the observer to the cluster.

  The rate equations for the photon distribution function $n(x)$ and the spectral intensity function $I(x)$ were derived in the NK paper\cite{noza09}, where $x=\omega/k_{B}T_{CMB}$.  We recall the results here to make the present paper more self-contained.  They are given as follows:
\begin{eqnarray}
&&\hspace{-10mm}
\frac{\partial n(x)}{\partial \tau}
 = \int_{-\infty}^{\infty}ds P_1(s,\beta_{c,z})
\left[n(e^sx)- n(x)\right] \, ,
\label{eq3a-1}   \\
&&\hspace{-10mm}
\frac{\partial I(x)}{\partial \tau}
 = \int_{-\infty}^{\infty}ds
{P}_1(s,\beta_{c,z}) \left[e^{-3s}I(e^{s}x)- I(x)\right]  \, ,
\label{eq3a-2}  \\
&&\hspace{+12mm}
d\tau  =  n_e\sigma_T dt \, , 
\label{eq3a-3} \\
&&\hspace{0mm}
P_1(s,\beta_{c,z}) = P_{1}(s) + \beta_{c,z} P_{1, K}(s)  \, ,
\label{eq3a-4}
\end{eqnarray}
where $\beta_{c,z}$ is the peculiar velocity of the CG parallel to the observer.  It should be noted that $O(\beta_{c,z}^{2})$ and higher-order contributions were neglected in deriving Eq.~(\ref{eq3a-4}), because $\beta_{c,z} \ll 1$ is satisfied for most of the CG.  The typical value is $\beta_{c,z}$=1/300 for $v_{c}$=1,000km/s.  In Eq.~(\ref{eq3a-4}), $P_{1}(s)$ is the redistribution function defined by Eq.~(\ref{eq2a-4}).

Similarly, in Eq.~(\ref{eq3a-4}) $P_{1,K}(s)$ is the redistribution function for photon of a frequency shift $s$ in the case of the non-zero peculiar velocity of the CG,
\begin{eqnarray}
&&\hspace{-10mm}
 P_{1,K}(s) = \int_{\beta_{min}}^{1}d\beta\beta^2\gamma^5 \tilde{p}_e(E)P_{K}(s,\beta)  \, ,
\label{eq3a-5}  \\
&&\hspace{-10mm}
P_{K}(s,\beta)
= \frac{e^{s}}{2\beta\gamma^4} \left(\frac{\gamma}{\theta_{e}} \right)
\int_{\mu_1(s)}^{\mu_2(s)}d\mu_0 (\beta \mu_{0}-\beta^2)
\nonumber  \\
&&\hspace{+27mm}
\times \frac{1}{(1-\beta\mu_0)^3}
f\left(\mu_0, \mu_{0}^{\prime} \right)   \, ,
\label{eq3a-6}
\end{eqnarray}
where $f\left(\mu_0, \mu_{0}^{\prime} \right)$ is given by Eq.~(\ref{eq2a-6}), and variables appearing in Eqs.~(\ref{eq3a-5}) and (\ref{eq3a-6}) are given by Eqs.~(\ref{eq2a-8}) -- (\ref{eq2a-11}).  It should be noted that Eq.~(\ref{eq3a-6}) was derived for electrons in thermal equilibrium at a temperature $T_{e}$.  The distribution function for electron of a velocity $\beta$ is given by
\begin{eqnarray}
&&\hspace{-10mm}
\tilde{p}_{e}(E) = \frac{1}{\theta_{e} K_{2}(1/\theta_{e})} e^{-\gamma/\theta_{e}}  \, ,
\label{eq3a-7}
\end{eqnarray}
where $\theta_{e}=k_{B}T_{e}/mc^{2}$, and $K_{2}(z)$ is the modified Bessel function of the second kind.  For the power-law distributions, $\gamma/\theta_{e}$ in Eq.~(\ref{eq3a-6}) should be replaced by $a/\beta^2$ and $a$ for the $p$-power distribution $\tilde{p}_{e}(E) \propto p^{-a}$ and the $E$-power distribution $\tilde{p}_{e}(E) \propto E^{-a}$, respectively.

\subsection{Properties of $P_{1}(s)$ and $P_{1,K}(s)$}

  In the present subsection, we study the properties of the redistribution functions $P_{1}(s)$ and $P_{1,K}(s)$.  It is familiar that $P_{1}(s)$ satisfies Eq.~(\ref{eq2a-7}).  We show that $P_{1,K}(s)$ also satisfies
\begin{eqnarray}
&&\hspace{-10mm}
\int_{-\infty}^{\infty}ds P_{1,K}(s) = 1 \, ,
\label{eq3b-1}
\end{eqnarray}
if electrons are in a thermal equilibrium state.  Integrating Eq.~(\ref{eq3a-5}) over $s$, one obtains the following:
\begin{eqnarray}
&&\hspace{-10mm}
\int_{-\infty}^{\infty}ds P_{1,K}(s) = \int_{0}^{\infty}d\tilde{p} \tilde{p}^{2} \tilde{p}_{e}(E) \left(\frac{\gamma}{\theta_{e}} \right) 
\nonumber  \\
&&\hspace{-1mm}
\times \frac{1}{2\gamma^4} \int_{-1}^{+1}d\mu_0 \frac{\beta \mu_{0}-\beta^2}{(1-\beta\mu_0)^4} 
\int_{-1}^{+1}d\mu_0^{\prime} f\left(\mu_0, \mu_{0}^{\prime} \right) \, ,
\label{eq3b-2}
\end{eqnarray}
where $\tilde{p} \equiv p/m$.  In deriving Eq.~(\ref{eq3b-2}), the integral variables $\beta$ and $s$ were replaced by $\tilde{p}$ and $\mu_{0}^{\prime}$, respectively.  Equation (\ref{eq3b-2}) is further simplified by inserting Eq.~({\ref{eq3a-7}) and using the following relations:
\begin{eqnarray}
&&\hspace{-10mm}
\int_{-1}^{+1}d\mu_0^{\prime} f\left(\mu_0, \mu_{0}^{\prime} \right) = 1 \, ,
\label{eq3b-3}  \\
&&\hspace{-20mm}
\frac{1}{2\gamma^4} \int_{-1}^{+1}d\mu_0 \frac{\beta \mu_{0}-\beta^2}{(1-\beta\mu_0)^4} = \frac{1}{3} \beta^{2} \, .
\label{eq3b-4}
\end{eqnarray}
One has
\begin{eqnarray}
&&\hspace{-10mm}
\int_{-\infty}^{\infty}ds P_{1,K}(s) = \int_{0}^{\infty}d\tilde{p} \tilde{p}^{2} \tilde{p}_{e}(E) \frac{1}{3} \beta^{2} \left(\frac{\gamma}{\theta_{e}} \right) \, ,
\nonumber  \\
&&\hspace{0mm}
= \frac{1}{\theta_{e} K_{2}(1/\theta_{e})} \int_{0}^{\infty}d\tilde{p} \tilde{p}^{3} \frac{1}{3 \theta_{e}}\left\{ (- \theta_{e}) e^{-\tilde{E}/\theta_{e}} \right\}^{\prime} \, ,
\label{eq3b-5}
\end{eqnarray}
where $\beta=\tilde{p}/\tilde{E}$, $\gamma=\tilde{E}$, and $\tilde{E} \equiv E/m$ were used.  In the second line of Eq.~(\ref{eq3b-5}), $\{ f(\tilde{p}) \}^{\prime}$ denotes the derivative of $f(\tilde{p})$ by $\tilde{p}$.  Applying the partial integration to Eq.~(\ref{eq3b-5}), one finally obtains
\begin{eqnarray}
&&\hspace{-10mm}
\int_{-\infty}^{\infty}ds P_{1,K}(s) 
=  \frac{1}{ \theta_{e} K_{2}(1/\theta_{e})} \left(-\frac{1}{3} \right) \left[ \tilde{p}^{3} e^{-\tilde{E}/\theta_{e}} \right]^{\infty}_{0} 
\nonumber  \\
&&\hspace{17mm}
+ \frac{1}{ \theta_{e} K_{2}(1/\theta_{e})} \int_{0}^{\infty}d\tilde{p} \tilde{p}^{2} e^{-\tilde{E}/\theta_{e}}  \, ,
\nonumber \\
&&\hspace{+14mm}
 = \, \, 1 \, .
\label{eq3b-6}
\end{eqnarray}
Thus, Eq.~(\ref{eq3b-1}) was shown.

\begin{figure}
\begin{center}
\includegraphics[angle=0,width=0.5\textwidth]{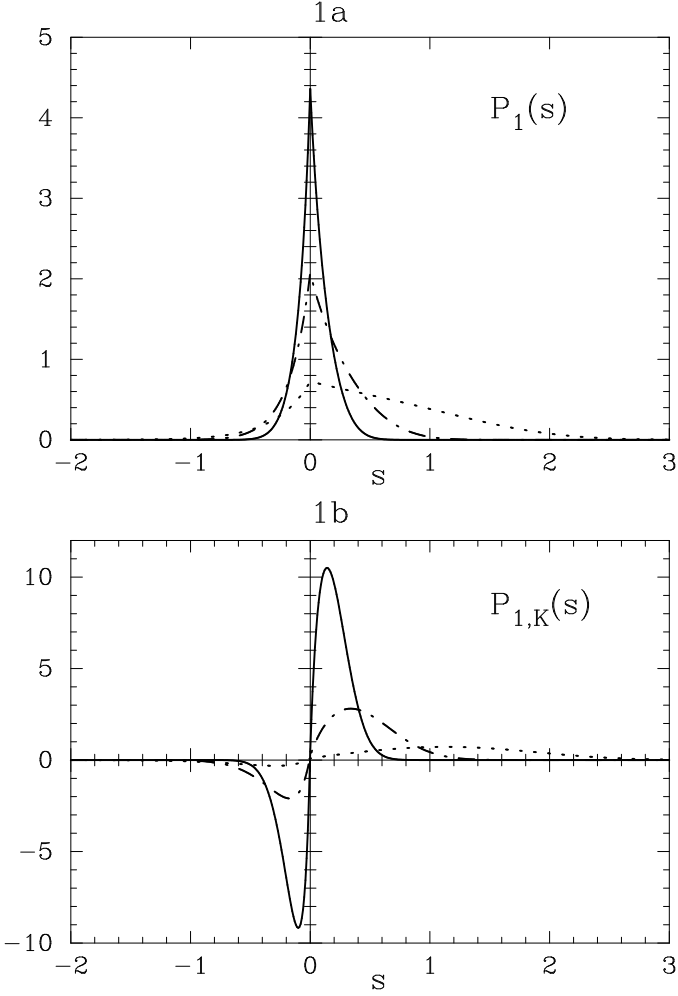}
\end{center}
\caption{Plotting of $P_{1}(s)$ and $P_{1,K}(s)$ as a function of $s$.  Figs.~1a and 1b are $P_{1}(s)$ and $P_{1,K}(s)$, respectively.  The solid curve, dash-dotted curve, and dotted curve correspond to $k_{B}T_{e}$ = 5keV, 20keV, and 100keV, respectively.}
\end{figure}

It is worthwhile to mention as follows:  As seen from the right-hand side of the first line of Eq.~(\ref{eq3b-5}), Eq.~(\ref{eq3b-5}) is proportional to the pressure of the electron distribution, because the pressure of the electron is described in general by
\begin{eqnarray}
&&\hspace{-10mm}
P = n_{e} mc \int_{0}^{\infty}d\tilde{p} \tilde{p}^{2} \tilde{p}_{e}(E) \frac{1}{3} \tilde{p} v  \, ,
\label{eq3b-7}
\end{eqnarray}
where $v=\beta c$.  For the thermal distribution, one has $P_{thermal}= n_{e} m c \theta_{e}$, which again leads to Eq.~(\ref{eq3b-1}).  For the cases of nonthermal distributions, in general, Eq.~(\ref{eq3b-1}) is not satisfied.  This will be discussed in Sec.~IV.

More importantly, the present finding suggests that one can interpret $P_{1,K}(s)$ as {\it the pressure distribution function} for a frequency shift $s$.  In the present paper, we call $P_{1,K}(s)$ the pressure distribution function hereafter.

In Figs.~1a and 1b, we plot the frequency redistribution function $P_{1}(s)$ and the pressure distribution function $P_{1,K}(s)$, respectively, as a function of $s$.  The $k_{B}T_{e}$ dependences are illustrated.  The solid curve, dash-dotted curve, and dotted curve correspond to $k_{B}T_{e}$ = 5keV, 20keV, and 100keV, respectively.  In Fig.~1a, the familiar behavior of the redistribution function $P_{1}(s)$ is shown, where the sharp peak at $s$=0 for $k_{B}T_{e}$=5keV becomes flatter and wider for higher electron temperatures.  On the other hand, $P_{1,K}(s)$ is shown for the first time.  It is seen from Fig.~1b that $P_{1,K}(s)$ is not positive definite, which is interpreted as the pressure distribution.  It is seen that $P_{1,K}(s)$ also has a large $k_{B}T_{e}$ dependence.  Two sharp peaks of opposite signs are canceling each other for the case $k_{B}T_{e}$=5keV, whereas the peaks become flatter and wider for higher electron temperatures.

\subsection{Formal solutions}

  Let us now derive the formal solutions of the rate equations of Eqs.~(\ref{eq3a-1}) and (\ref{eq3a-2}) in the operator representation.  First, introduce the operators $\mathcal{O}(-D)$ and $\mathcal{O}_{K}(D-n)$ by
\begin{eqnarray}
&&\hspace{-10mm}
\mathcal{O}(-D) = \int_{-\infty}^{\infty}ds P_1(s) e^{-sD} - 1 \, ,
\label{eq3c-1}  \\
&&\hspace{-10mm}
\mathcal{O}_{K}(D-n) = \int_{-\infty}^{\infty}ds P_{1,K}(s) 
\left[ e^{s(D-n)} - 1 \right]  \, ,
\label{eq3c-2}
\end{eqnarray}
where $n$ is an integer.  Applying Eq.~(\ref{eq3c-1}) and Eq.~(\ref{eq3c-2}) of $n=3$ to the spectral intensity function $I(x)$, one obtains
\begin{eqnarray}
&&\hspace{-10mm}
\mathcal{O}(-D) I(x) = \int_{-\infty}^{\infty}ds P_1(s) \left[ I(e^{-s}x) - I(x) \right] \, ,
\label{eq3c-3}  \\
&&\hspace{-10mm}
\mathcal{O}_{K}(D-3) I(x) = \int_{-\infty}^{\infty}ds P_{1,K}(s) 
\nonumber  \\
&&\hspace{+20mm}
\times \left[ e^{-3s} I(e^{s}x) - I(x) \right] \, .
\label{eq3c-4}
\end{eqnarray}
Note that, in the case of the thermal distribution, Eq.~(\ref{eq3c-2}) is further simplified as
\begin{eqnarray}
&&\hspace{-10mm}
\mathcal{O}_{K}(D-n) = \int_{-\infty}^{\infty}ds P_{1,K}(s) e^{s(D-n)} - 1 \, ,
\label{eq3c-5}
\end{eqnarray}
because Eq.~(\ref{eq3b-1}) is valid.  The rate equation of Eq.~(\ref{eq3a-2}) is rewritten  as follows:
\begin{eqnarray}
&&\hspace{-10mm}
\frac{\partial I(x)}{\partial \tau}
 = \left[ \mathcal{O}(-D) + \beta_{c,z} \mathcal{O}_{K}(D-3) \right] I(x)  \, .
\label{eq3c-6}
\end{eqnarray}

Now we expand the solution $I(x)$ in powers of $\beta_{c,z}$ as follows:
\begin{eqnarray}
&&\hspace{-10mm}
I(x) = I_{SZ}(x) + \beta_{c,z} I_{kSZ}(x) + O(\beta_{c,z}^{2})  \, ,
\label{eq3c-7}
\end{eqnarray}
where $SZ$ and $kSZ$ denote the SZ effect and kinematical SZ effect of the first-order in $\beta_{c,z}$, respectively.  Inserting Eq.~(\ref{eq3c-7}) into Eq.~(\ref{eq3c-6}), one obtains the rate equation for each-order in $\beta_{c,z}$,
\begin{eqnarray}
&&\hspace{-10mm}
\frac{\partial I_{SZ}(x)}{\partial \tau}
 = \mathcal{O}(-D) I_{SZ}(x)   \, ,
\label{eq3c-8}  \\
&&\hspace{-10mm}
\frac{\partial I_{kSZ}(x)}{\partial \tau}
 = \mathcal{O}(-D) I_{kSZ}(x) + g(x, \tau)  \, ,
\label{eq3c-9}
\end{eqnarray}
where
\begin{eqnarray}
&&\hspace{-5mm}
g(x, \tau) =  \mathcal{O}_{K}(D-3) I_{SZ}(x)  \, ,
\nonumber  \\
&&\hspace{+6mm}
=  \mathcal{O}_{K}(D-3) e^{\tau \mathcal{O}(-D)} I_{0}(x)  \, .
\label{eq3c-10}
\end{eqnarray}
Equation (\ref{eq3c-8}) corresponds to the rate equation for the SZ effect, which is identical to Eq.~(\ref{eq2b-10}).  Therefore, the formal solution is given by Eq.~(\ref{eq2b-12}).  On the other hand, Eq.~(\ref{eq3c-9}) corresponds to the rate equation for the kinematical SZ effect.

  Let us now solve Eq.~({\ref{eq3c-9}) with an initial condition $I_{kSZ}(x)$=0 at $\tau$=0.  Introducing a new function $u(x)$ by
\begin{eqnarray}
&&\hspace{-10mm}
I_{kSZ}(x) = e^{\tau \mathcal{O}(-D)} u(x) \, ,
\label{eq3c-11}
\end{eqnarray}
and inserting it into Eq.~(\ref{eq3c-9}), one has the equation for $u(x)$ as follows:
\begin{eqnarray}
&&\hspace{-10mm}
\frac{\partial u(x)}{\partial \tau}
 = e^{-\tau \mathcal{O}(-D)} g(x, \tau)  \, .
\label{eq3c-12}
\end{eqnarray}
The solution is
\begin{eqnarray}
&&\hspace{-10mm}
u(x) = \int_{0}^{\tau}d \tau^{\prime} e^{-\tau^{\prime} \mathcal{O}(-D)} g(x, \tau^{\prime})   \, ,
\label{eq3c-13}
\end{eqnarray}
where the initial condition $u(x)$=0 at $\tau$=0 was used.  Thus, one has
\begin{eqnarray}
&&\hspace{-10mm}
I_{kSZ}(x) = \int_{0}^{\tau}d \tau^{\prime} e^{( \tau -\tau^{\prime}) \mathcal{O}(-D)} g(x, \tau^{\prime}) \, ,
\nonumber  \\
&&\hspace{+3mm}
 = \int_{0}^{\tau}d \tau^{\prime} e^{( \tau -\tau^{\prime}) \mathcal{O}(-D)} \mathcal{O}_{K}(D-3) \,
\nonumber  \\
&&\hspace{+15mm}
 \times e^{\tau^{\prime} \mathcal{O}(-D)} I_{0}(x) \, .
\label{eq3c-14}
\end{eqnarray}
Equation (\ref{eq3c-14}) is further simplified since the operators $\mathcal{O}(-D)$ and $\mathcal{O}_{K}(D-3)$ are commutable each other.  Finally, one obtains the following formal solution for the spectral intensity function of the kinematical SZ effect:
\begin{eqnarray}
&&\hspace{-10mm}
I_{kSZ}(x) = \tau \mathcal{O}_{K}(D-3) e^{ \tau \mathcal{O}(-D)} I_{0}(x) \, ,
\nonumber  \\
&&\hspace{+3mm}
= \tau \mathcal{O}_{K}(D-3) I_{SZ}(x)  \, ,
\nonumber  \\
&&\hspace{+5mm}
= \tau g(x, \tau)  \, .
\label{eq3c-15add}
\end{eqnarray}

  Similarly, the rate equation for the photon distribution function $n(x)$ is expressed by
\begin{eqnarray}
&&\hspace{-10mm}
\frac{\partial n(x)}{\partial \tau}
 = \left[ \mathcal{O}(D) + \beta_{c,z} \mathcal{O}_{K}(D) \right] n(x)  \, .
\label{eq3c-15}
\end{eqnarray}
The initial condition is $n(x)=n_{0}(x)$ at $\tau$=0.  Expanding $n(x)$ in powers of $\beta_{c,z}$ by
\begin{eqnarray}
&&\hspace{-10mm}
n(x) = n_{SZ}(x) + \beta_{c,z} n_{kSZ}(x) + O(\beta_{c,z}^{2})  \, ,
\label{eq3c-16}
\end{eqnarray}
one finally obtains the formal solutions for the SZ effect and kinematical SZ effect as follows:
\begin{eqnarray}
&&\hspace{-10mm}
n_{SZ}(x) = e^{\tau \mathcal{O}(D)} n_{0}(x)  \, ,
\label{eq3c-17}  \\
&&\hspace{-10mm}
n_{kSZ}(x) = \tau \mathcal{O}_{K}(D) n_{SZ}(x)  \, .
\label{eq3c-18}
\end{eqnarray}
In deriving Eqs.~(\ref{eq3c-17}) and (\ref{eq3c-18}), the initial conditions $n_{SZ}(x)=n_{0}(x)$ and $n_{kSZ}(x)$=0 at $\tau$=0 were used.  Before closing this section, it should be noted that Eqs.~(\ref{eq3c-15add}) and (\ref{eq3c-18}) require the following relation for $\mathcal{O}_{K}$:
\begin{eqnarray}
&&\hspace{-10mm}
x^3 \mathcal{O}_{K}(D) = \mathcal{O}_{K}(D-3) x^{3}  \, ,
\label{eq3c-19}
\end{eqnarray}
which is satisfied by Eq.~(\ref{eq3c-2}).

\section{Numerical solutions}

\subsection{Derivation of numerical solutions}

In this section we show the derivation of the numerical solutions for the rate equations of Eqs.~(\ref{eq3c-8}) and (\ref{eq3c-9}).  We consider an ideal condition that the CG is infinitely large.  The rate equations are expressed as follows:
\begin{eqnarray}
&&\hspace{-10mm}
\frac{\partial I_{SZ}(x)}{\partial \tau}
= \int_{-\infty}^{\infty}ds P_1(s) \left[ I_{SZ}(e^{-s}x) - I_{SZ}(x) \right] \, ,
\label{eq4a-1}  \\
&&\hspace{-10mm}
\frac{\partial I_{kSZ}(x)}{\partial \tau}
= \int_{-\infty}^{\infty}ds P_1(s) \left[ I_{kSZ}(e^{-s}x) - I_{kSZ}(x) \right]
\nonumber  \\
&&\hspace{15mm}
+ \, g(x, \tau)   \, ,
\label{eq4a-2}
\end{eqnarray}
where $g(x, \tau)$ is defined by Eq.~(\ref{eq3c-10}).  The explicit form is
\begin{eqnarray}
&&\hspace{-10mm}
g(x, \tau) = \int_{-\infty}^{\infty}ds {P}_{1,K}(s) \left[e^{-3s}I_{SZ}(e^{s}x)- I_{SZ}(x) \right]   \, .
\label{eq4a-2.5}
\end{eqnarray}
The initial conditions at $\tau=0$ are given by
\begin{eqnarray}
&&\hspace{-10mm}
 I_{SZ}(x) = I_{0}(x) = I_{0} \frac{x^{3}}{e^{x}-1}  \, ,
\label{eq4a-3}  \\
&&\hspace{-10mm}
 I_{kSZ}(x) = 0 \, ,
\label{eq4a-4}  \\
&&\hspace{-10mm}
g(x,0) = \int_{-\infty}^{\infty}ds {P}_{1,K}(s) \left[e^{-3s}I_{0}(e^{s}x)- I_{0}(x) \right]    \, ,
\label{eq4a-5}
\end{eqnarray}
where $I_{0}=(k_{B}T_{CMB})^{3}/2\pi^{2}$.

Let us first solve Eq.~(\ref{eq4a-1}).  Introducing $\tilde{I}_{SZ}(x, \tau)$ by
\begin{eqnarray}
&&\hspace{-10mm}
I_{SZ}(x) = e^{-\tau} \tilde{I}_{SZ}(x, \tau)  \, ,
\label{eq4a-6}
\end{eqnarray}
and inserting it into Eq.~(\ref{eq4a-1}), one obtains the following equation for $\tilde{I}_{SZ}(x, \tau)$:
\begin{eqnarray}
&&\hspace{-10mm}
\frac{\partial \tilde{I}_{SZ}(x, \tau)}{\partial \tau}
= \int_{-\infty}^{\infty}ds P_1(s) \tilde{I}_{SZ}(e^{-s}x, \tau) \, .
\label{eq4a-7}
\end{eqnarray}
Integrating Eq.~(\ref{eq4a-7}) between $\tau$ and $\tau+\Delta \tau$ ($\Delta \tau \ll 1$), and applying the Runge-Kutta method of the fourth-order, one obtains the following difference equation:
\begin{eqnarray}
&&\hspace{-10mm}
\tilde{I}_{SZ}(x, \tau+\Delta \tau) = \tilde{I}_{SZ}(x, \tau)
\nonumber \\
&&\hspace{+10mm}
 + a(\Delta \tau) 
 \int_{-\infty}^{\infty}ds P_1(s) \tilde{I}_{SZ}(e^{-s}x, \tau) \, ,
\label{eq4a-8}  \\
&&\hspace{-10mm}
a(\Delta \tau) = \Delta \tau + \frac{1}{2!} (\Delta \tau)^{2} + \frac{1}{3!} (\Delta \tau)^{3} + \frac{1}{4!} (\Delta \tau)^{4} \, .
\label{eq4a-9}
\end{eqnarray}

(i) Inserting $\tau=0$ into Eq.~(\ref{eq4a-8}), one has the solution at $\tau= \Delta \tau$,
\begin{eqnarray}
&&\hspace{-12mm}
\tilde{I}_{SZ}(x, \Delta \tau) = I_{0}(x) + a(\Delta \tau)
 \int_{-\infty}^{\infty}ds P_1(s) I_{0}(e^{-s}x) \, ,
\label{eq4a-10}
\end{eqnarray}
where the initial condition of Eq.~(\ref{eq4a-3}) was used.  In Eq.~(\ref{eq4a-10}), the right-hand side (RHS) can be calculated with the initial function of Eq.~(\ref{eq4a-3}).  (ii) Similarly, inserting $\tau=\Delta \tau$ into Eq.~(\ref{eq4a-8}), one obtains the solution at $\tau= 2\Delta \tau$,
\begin{eqnarray}
&&\hspace{-10mm}
\tilde{I}_{SZ}(x, 2\Delta \tau) = \tilde{I}_{SZ}(x, \Delta \tau)
\nonumber  \\
&&\hspace{+10mm}
 + a(\Delta \tau) 
 \int_{-\infty}^{\infty}ds P_1(s) \tilde{I}_{SZ}(e^{-s}x, \Delta \tau) \, ,
\label{eq4a-11}
\end{eqnarray}
where the RHS can be calculated with Eq.~(\ref{eq4a-10}).  (iii) Repeating the same step $n$-times, one obtains the solution at $\tau=n \Delta \tau$,
\begin{eqnarray}
&&\hspace{-10mm}
\tilde{I}_{SZ}(x, n\Delta \tau) = \tilde{I}_{SZ}(x, (n-1)\Delta \tau)
\nonumber  \\
&&\hspace{0mm}
 + a(\Delta \tau) 
\int_{-\infty}^{\infty}ds P_1(s) \tilde{I}_{SZ}(e^{-s}x, (n-1)\Delta \tau) \, ,
\label{eq4a-12}
\end{eqnarray}
where the RHS can be calculated with the solution at $\tau=(n-1)\Delta \tau$.  Thus, the numerical solution $I_{SZ}(x)$ at $\tau=N\Delta \tau$ has been obtained.  We call this method the full-order numerical calculation in contrast to the first-order numerical calculation in the $\tau$-expansion assuming $\tau \ll1$, which was done by Itoh, Kohyama, and Nozawa\cite{itoh98}.

Similarly, Eq.~(\ref{eq4a-2}) is also solved numerically.  Introducing $\tilde{I}_{kSZ}(x, \tau)$ by
\begin{eqnarray}
&&\hspace{-10mm}
I_{kSZ}(x) = e^{-\tau} \tilde{I}_{kSZ}(x, \tau)  \, ,
\label{eq4a-13}
\end{eqnarray}
and inserting it into Eq.~(\ref{eq4a-2}), one obtains the following equation for $\tilde{I}_{kSZ}(x, \tau)$:
\begin{eqnarray}
&&\hspace{-12mm}
\frac{\partial \tilde{I}_{kSZ}(x, \tau)}{\partial \tau}
= \int_{-\infty}^{\infty}ds P_1(s) \tilde{I}_{kSZ}(e^{-s}x, \tau) + \tilde{g}(x, \tau) \, ,
\label{eq4a-14}  \\
&&\hspace{-4mm}
\tilde{g}(x, \tau) = \int_{-\infty}^{\infty}ds {P}_{1,K}(s) \left[e^{-3s}\tilde{I}_{SZ}(e^{s}x, \tau) \right.
\nonumber  \\
&&\hspace{+43mm}
\left. - \tilde{I}_{SZ}(x, \tau) \right]    \, .
\label{eq4a-15} 
\end{eqnarray}
It should be noted that $\tilde{g}(x, \tau)$ in Eq.~(\ref{eq4a-14}) is a known function, because $\tilde{I}_{SZ}(x, \tau)$ in Eq.~(\ref{eq4a-15}) was already obtained by solving Eq.~(\ref{eq4a-7}).  Integrating Eq.~(\ref{eq4a-14}) between $\tau$ and $\tau+\Delta \tau (\Delta \tau \ll 1)$, and applying the Runge-Kutta method of the fourth-order, one obtains the following difference equation:
\begin{eqnarray}
&&\hspace{-10mm}
\tilde{I}_{kSZ}(x, \tau+\Delta \tau) = \tilde{I}_{kSZ}(x, \tau)
\nonumber \\
&&\hspace{10mm}
 + a(\Delta \tau) \left\{
 \int_{-\infty}^{\infty}ds P_1(s) \tilde{I}_{kSZ}(e^{-s}x, \tau) \right.
\nonumber  \\
&&\hspace{45mm}
\left. + \, \tilde{g}(x, \tau) \biggr\} \right. \, .
\label{eq4a-16}
\end{eqnarray}

(i) Inserting $\tau=0$ into Eq.~(\ref{eq4a-16}), one obtains the solution at $\tau=\Delta \tau$,
\begin{eqnarray}
&&\hspace{-10mm}
\tilde{I}_{kSZ}(x, \Delta \tau) = a(\Delta \tau) g(x, 0) \, ,
\label{eq4a-17}
\end{eqnarray}
where the initial conditions of Eqs.~(\ref{eq4a-4}) and (\ref{eq4a-5}) were used.  (ii) Similarly, inserting $\tau=\Delta \tau$ into Eq.~(\ref{eq4a-16}), one obtains the solution at $\tau= 2\Delta \tau$,
\begin{eqnarray}
&&\hspace{-10mm}
\tilde{I}_{kSZ}(x, 2\Delta \tau) = \tilde{I}_{kSZ}(x, \Delta \tau)
\nonumber  \\
&&\hspace{0mm}
 + a(\Delta \tau) \left\{
 \int_{-\infty}^{\infty}ds P_1(s) \tilde{I}_{kSZ}(e^{-s}x, \Delta \tau) 
 \right.
\nonumber  \\
&&\hspace{40mm}
\left. + \, \tilde{g}(x, \Delta \tau) \biggr\} \right.  \, ,
\label{eq4a-18}
\end{eqnarray}
where the RHS is calculated with Eq.~(\ref{eq4a-17}).  (iii) Repeating the same step $n$-times, one obtains the solution at $\tau=n \Delta \tau$,
\begin{eqnarray}
&&\hspace{-10mm}
\tilde{I}_{kSZ}(x, n\Delta \tau) = \tilde{I}_{kSZ}(x, (n-1)\Delta \tau)
\nonumber  \\
&&\hspace{-5mm}
 + a(\Delta \tau) \left\{
 \int_{-\infty}^{\infty}ds P_1(s) \tilde{I}_{kSZ}(e^{-s}x, (n-1) \Delta \tau) \right.
\nonumber  \\
&&\hspace{33mm}
\left. + \, \tilde{g}(x, (n-1) \Delta \tau) \biggr\}  \right. \, ,
\label{eq4a-19}
\end{eqnarray}
where the RHS can be calculated with the solution at $\tau=(n-1)\Delta \tau$.  Thus, the full-order numerical solution $I_{kSZ}(x)$ at $\tau=n\Delta \tau$ has been obtained.  It should be noted that $I_{kSZ}(x)$ and $n_{kSZ}(x)$ can be also calculated directly from Eqs.~(\ref{eq3c-15add}) and (\ref{eq3c-18}), respectively.

\subsection{Results for thermal distribution}

  In this section we present the results in the full-order numerical calculations of $I_{SZ}(x)$ and $I_{kSZ}(x)$ for the thermal electron distribution.  First we define the spectral distortion functions as follows:
\begin{eqnarray}
&&\hspace{-10mm}
\Delta I_{SZ}(x)/(\tau I_{0}) = \left(I_{SZ}(x) - I_{0}(x) \right)/(\tau I_{0})   \, ,
\label{eq4b-1}  \\
&&\hspace{-10mm}
\Delta I_{kSZ}(x)/(\tau I_{0}) = I_{kSZ}(x)/(\tau I_{0})  \, ,
\label{eq4b-2}
\end{eqnarray}
where $I_{0}(x)$ is defined by Eq.~(\ref{eq4a-3}).

\begin{figure}
\begin{center}
\includegraphics[angle=0,width=0.5\textwidth]{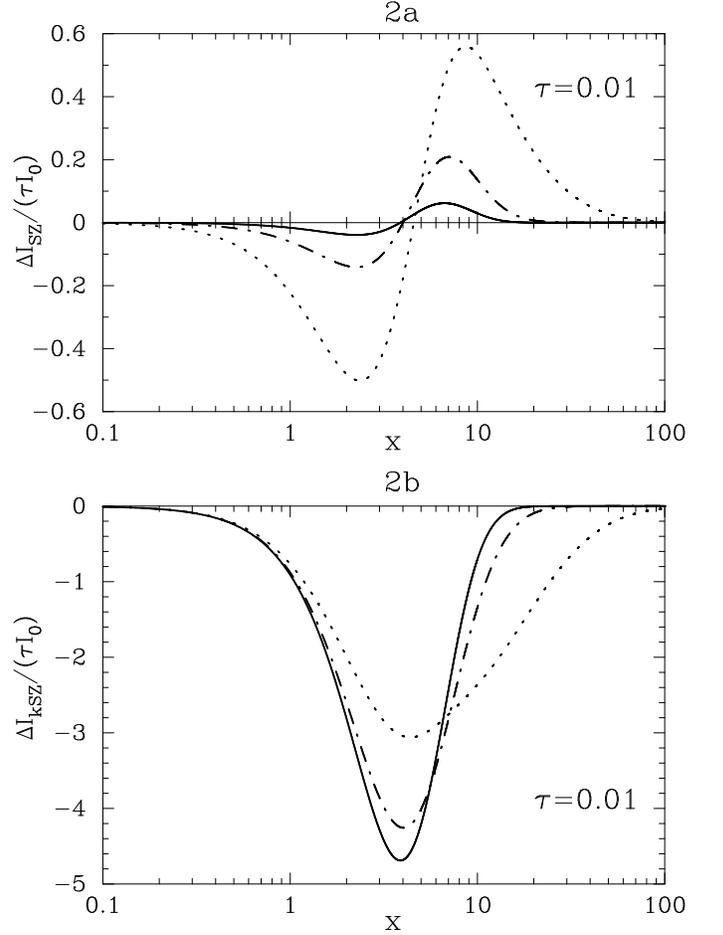}
\end{center}
\caption{Plotting of spectral distortion functions for $\tau=0.01$.  Figs.~2a and 2b are $\Delta I_{SZ}(x)$ and $\Delta I_{kSZ}(x)$, respectively.  The solid curve, dash-dotted curve, and dotted curve correspond to $k_{B}T_{e}$ = 5keV, 20keV, and 100keV, respectively.}
\end{figure}

We plot the spectral distortion functions as a function of $x$ in Fig.~2.  In Figs.~2a and 2b, the $k_{B}T_{e}$ dependences for $\Delta I_{SZ}(x)$ and $\Delta I_{kSZ}(x)$ are studied for the case of $\tau=0.01$.  The solid curve, dash-dotted curve, and dotted curve correspond to the full-order calculations for $k_{B}T_{e}$=5keV, 20keV, and 100keV, respectively.  For most of the CG, $k_{B}T_{e} \leq$20keV is satisfied, however, high $k_{B}T_{e}$ is still needed when analyzing high temperature clusters,  for example, Hansen, Pastor, and Semikoz\cite{hans02} showed that $k_{B}T_{e} \geq$100keV is needed in order to get 2$\sigma$ contraint on the temperature.  As seen from Fig.~2a, the peaks of $\Delta I_{SZ}(x)$ become higher and broader as $k_{B}T_{e}$ grows.  In Fig.~2a, we also plot the results of the first-order numerical calculations by Itoh, Kohyama, and Nozawa\cite{itoh98}.  The two curves are indistinguishable in the entire region of $x$.  This result guarantees that the first-order numerical calculation is valid for $\tau \ll 1$, which is satisfied for most of the CG.  Similarly, in Fig.~2b the peaks of $\Delta I_{kSZ}(x)$ become lower and broader as $k_{B}T_{e}$ grows.  In Fig.~2b, we also plot the results of the first-order numerical calculations, where the two curves are again indistinguishable in the entire region of $x$.  This result guarantees that the first-order numerical calculation is valid for the kinematical SZ effect of the CG which satisfies $\tau \ll 1$.  It is needless to mention that the signal of the kinematical SZ effect should be multiplied by the factor $\beta_{c,z}$.

\begin{figure}
\begin{center}
\includegraphics[angle=0,width=0.5\textwidth]{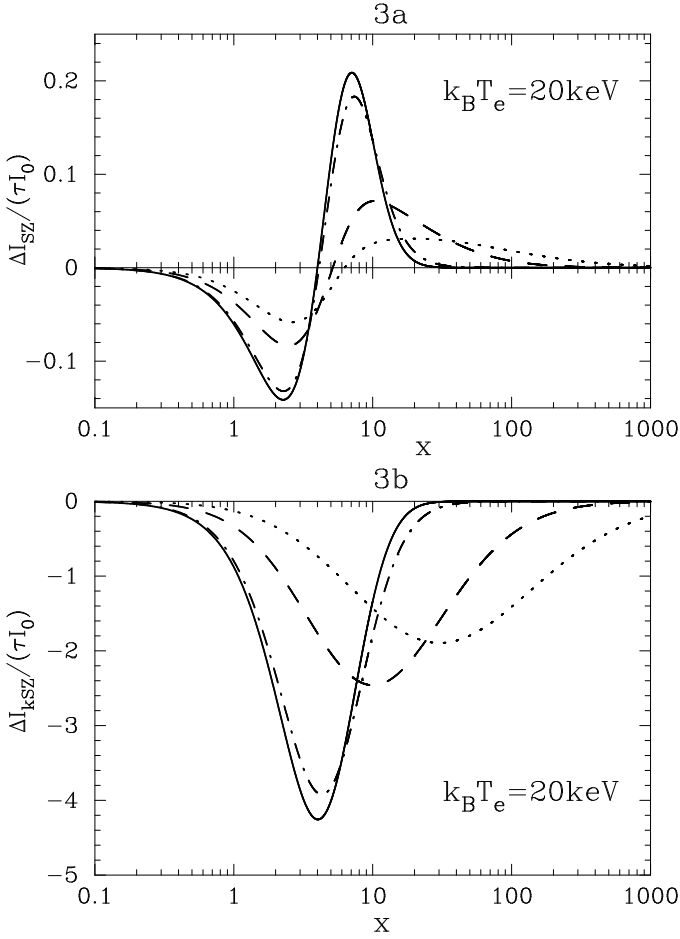}
\end{center}
\caption{Plotting of spectral distortion functions for $k_{B}T_{e}$=20keV.  Figs.~3a and 3b are $\Delta I_{SZ}(x)$ and $\Delta I_{kSZ}(x)$, respectively.  The solid curve, dash-dotted curve, dashed curve, and dotted curve correspond to $\tau$ = 0.01, 1, 10 and 20, respectively.}
\end{figure}

  In Figs.~3a and 3b, we show the $\tau$-dependence of the spectral distortion functions as a function of $x$ for $k_{B}T_{e}$=20keV.  The solid curve, dash-dotted curve, dashed curve, and dotted curve correspond to $\tau$=0.01, 1, 10, and 20, respectively.  For most of the CG, $\tau \ll 1$ is safely satisfied.  On the other hand, $\tau \gg 1$ cases are very important for optically thick plasmas, for example, in the accretion disk of the neutron star X-ray binaries\cite{suny85}, and references are therein.  It can be seen from Figs.~3a and 3b that the $\tau$-dependence of the redistribution functions is large, where the present full-order calculations are indispensable for the cases $\tau > 1$.  It can be seen that the peaks of $\Delta I_{SZ}(x)$ and $\Delta I_{kSZ}(x)$ become lower and extremely broader as $\tau$ grows.  The full-order calculation for $I_{SZ}(x)$ was also done by Dolgov et al.\cite{dolg01}.  The present calculation agrees with their calculation up to $x$=1,000.

  In the present calculation, we have checked the convergence of the numerical solution for the step-size $\Delta \tau$.  It has been found that $\Delta \tau$ = 10$^{-3}$ or less is necessary to get a good convergence, where the error was less than 0.05\% compared with other numerical calculation.  It was shown by Dolgov et al\cite{dolg01} that the solution has to be solved up to large values of $x$ for the cases $\tau \gg 1$.  This can be also seen from Figs.~3a and 3b.  In order to study the convergence, we have calculated $\Delta I_{SZ}(x)$ and $\Delta I_{kSZ}(x)$ up to $x$=50,000 for $\tau$=20.  The obtained ratios to the peak value at $x$=1,000 and $x$=10,000 for $\Delta I_{SZ}(x)$ are 5.75\% and 0.15\%, respectively.  Similarly, the ratios to the peak value at $x$=1,000 and $x$=10,000 for $\Delta I_{kSZ}(x)$ are 10.3\% and 0.36\%, respectively.

  Here, one should mention the difference between the present formalism and that of Dolgov et al.\cite{dolg01} in the limit $\tau \rightarrow \infty$.  The Boltzmann equation in Dolgov et al. has, as a stationary solution, the thermal equilibrium one of the temperature $T_{e}$.  On the other hand, the rate equations of Eqs.~(\ref{eq2a-1}) and (\ref{eq2a-2}) which are derived in the approximation $(\omega^{\prime}-\omega)/T_{e} \ll 1$ do not have the equilibrium solutions.  For the equilibrium state, for example, of $k_{B}T_{e} \sim$ 20keV, the above condition is no longer valid.  Thus, these equations are not applicable to the limit $\tau \rightarrow \infty$.  The equilibrium solutions are recovered in the present formalism by replacing Eqs.~(\ref{eq2a-1}) and (\ref{eq2a-2}) with new solutions $n(x,\eta)$ and $I(x,\eta)$ in the limit $\tau \rightarrow \infty$ as follows:
\begin{eqnarray}
&&\hspace{-10mm}
\frac{\partial n(x,\eta)}{\partial \tau}
 = \int_{-\infty}^{\infty}ds P_1(s)
\left[ n(e^sx,\eta) e^{\eta(e^{s}-1)x} \right.
\nonumber  \\
&&\hspace{+43mm}
 - n(x,\eta) \Bigr] \, ,
\label{eq4b-3}  \\
&&\hspace{-10mm}
\frac{\partial I(x,\eta)}{\partial \tau}
 = \int_{-\infty}^{\infty}ds P_1(s)
\left[ I(e^{-s}x,\eta) e^{\eta(e^{-s}-1)x} \right.
\nonumber  \\
&&\hspace{+43mm}
 - I(x,\eta) \Bigr] \, ,
\label{eq4b-4}
\end{eqnarray}
where $\eta \equiv T_{CMB}/T_{e}$.  In deriving Eqs.~(\ref{eq4b-3}) and (\ref{eq4b-4}), $n(x,\eta) \ll 1$ was assumed in the original Boltzmann equation.  In the equilibrium state, this condition is justified, because the chemical potential $|\mu| \gg 1$.  It can be seen that these equations take into account of the detailed balance corrections to Eqs.~(\ref{eq2a-1}) and (\ref{eq2a-2}).  The stationary solutions are
\begin{eqnarray}
&&\hspace{-10mm}
n(x,\eta) = e^{-\eta x + \mu}  \, ,
\label{eq4b-5}  \\
&&\hspace{-10mm}
I(x,\eta) = I_{0} x^{3} n(x,\eta) \, ,
\label{eq4b-6}
\end{eqnarray}
where $\mu$ = ln$\zeta$(3) + 3ln$\eta$ is the chemical potential in units of $k_{B}T_{e}$.  The cross over frequency $x_{c}$ in this limit is given by $x_{c}=-\mu$, where $\eta=1.174\times 10^{-8}$ and $\mu=-54.60$ at $k_{B}T_{e}$=20keV.

  Finally, we derive the formal solutions for the improved rate equations.  Introducing $n(x,\eta)=e^{-\eta x}\tilde{n}(x)$ and $I(x,\eta)=e^{-\eta x} \tilde{I}(x)$, Eqs.(\ref{eq4b-3}) and (\ref{eq4b-4}) are rewritten as follows:
\begin{eqnarray}
&&\hspace{-10mm}
\frac{\partial \tilde{n}(x)}{\partial \tau}
 = \int_{-\infty}^{\infty}ds P_1(s)
\left[ \tilde{n}(e^sx) - \tilde{n}(x) \right] \, ,
\label{eq4b-7}  \\
&&\hspace{-10mm}
\frac{\partial \tilde{I}(x)}{\partial \tau}
 = \int_{-\infty}^{\infty}ds P_1(s)
\left[ \tilde{I}(e^{-s}x) - \tilde{I}(x) \right] \, ,
\label{eq4b-8}
\end{eqnarray}
where $\tilde{n}(x)$ and $\tilde{I}(x)$ satisfy the same rate equations as Eqs.~(\ref{eq2a-1}) and (\ref{eq2a-2}), respectively.  Thus, the formal solutions are
\begin{eqnarray}
&&\hspace{-10mm}
n(x,\eta) = e^{-\eta x}  e^{\tau \mathcal{O}(D)} n_{0}(x)  \, ,
\label{eq4b-9} \\
&&\hspace{-10mm}
I(x,\eta) = e^{-\eta x}  e^{\tau \mathcal{O}(-D)} I_{0}(x) \, .
\label{eq4b-10}
\end{eqnarray}
where $n_{0}(x)$ and $I_{0}(x)$ are the initial solutions at $\tau=0$.

\begin{figure}
\begin{center}
\includegraphics[angle=0,width=0.5\textwidth]{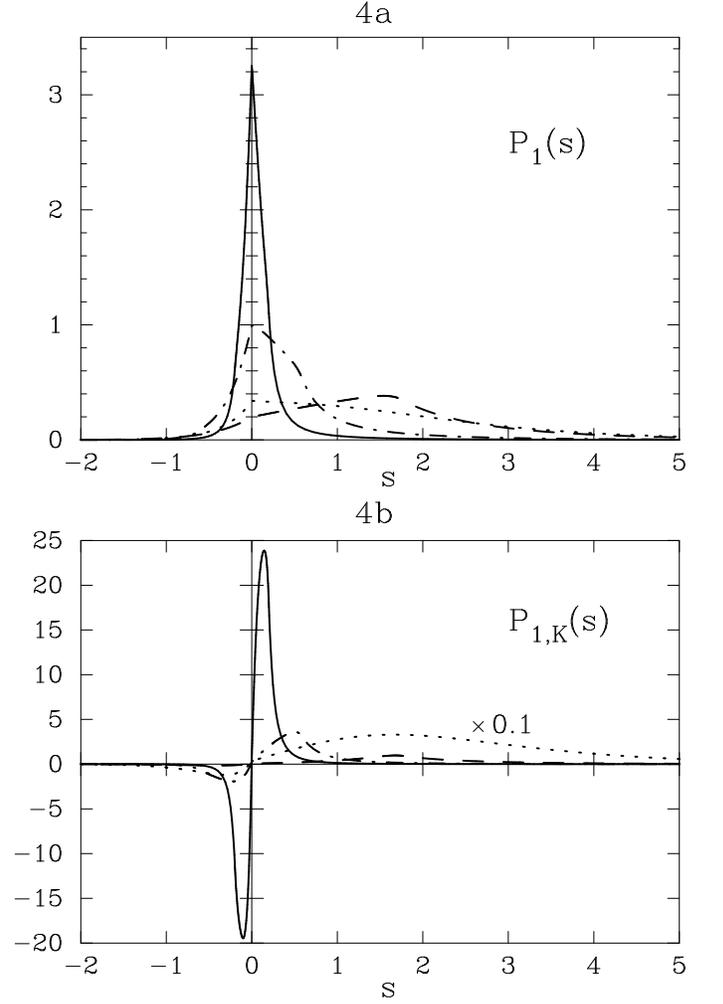}
\end{center}
\caption{Plotting of $P_{1}(s)$ and $P_{1,K}(s)$ for the nonthermal distributions.  Figs.~4a and 4b are $P_{1}(s)$ and $P_{1,K}(s)$, respectively.  The solid curve, dash-dotted curve, and dashed curve correspond to the $p$-power distributions for $\tilde{p}_{1}$ = 0.1, 0.3, and 1.0, respectively.  The dotted curve is the $E$-power distribution.  In Fig.~4b, the dotted curve is multiplied by a factor 10 in order to be visible in the same figure.}
\end{figure}

\begin{figure}
\begin{center}
\includegraphics[angle=0,width=0.5\textwidth]{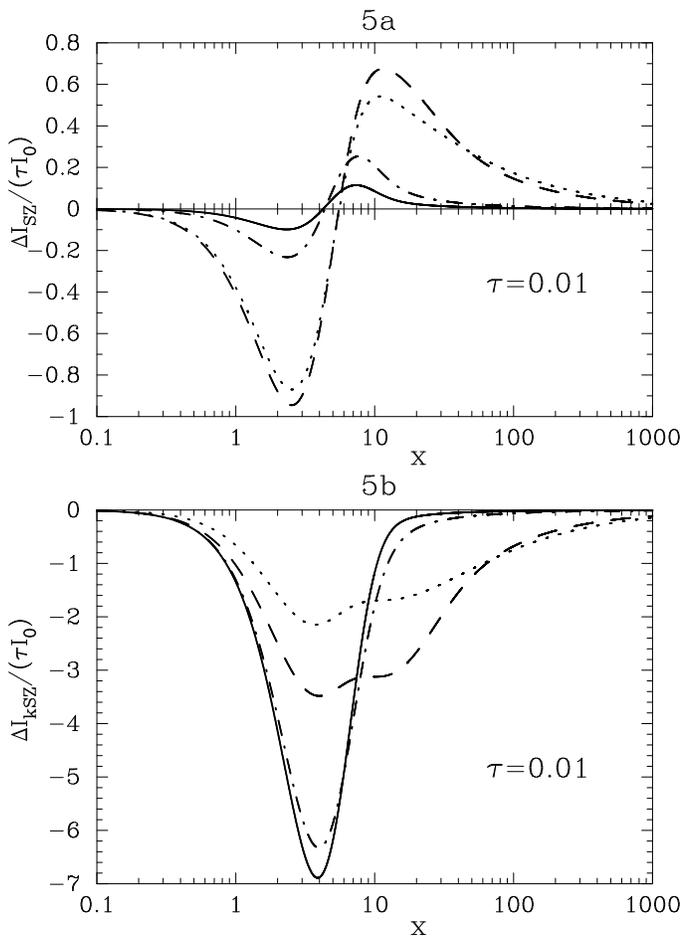}
\end{center}
\caption{Plotting of spectral distortion functions for the nonthermal distributions at $\tau$=0.01.  Figs.~5a and 5b are $\Delta I_{SZ}(x)$ and $\Delta I_{kSZ}(x)$, respectively.  The solid curve, dash-dotted curve, and dashed curve correspond to the $p$-power distributions for $\tilde{p}_{1}$ = 0.1, 0.3, and 1.0, respectively.  The dotted curve is the $E$-power distribution.}
\end{figure}

\subsection{Results for nonthermal distributions}

  In this section, we calculate the spectral distortion functions for the nonthermal electron distribution functions.  The careful study of the SZ effect on the nonthermal distributions was done, for example, by Colafrancesco et al.\cite{cola03, cola09}.  In the present paper, we study two types of single power-law distributions for illustrative purposes. \\
\noindent
(i) the $p$-power distribution:
\begin{eqnarray}
&&\hspace{-10mm}
\tilde{p}_{e}(E) = \left\{
\begin{array}{ll}
 A_{p} \, \tilde{p}^{-a_{p}} \, , &  \, \, \, \tilde{p}_{1} \leq \tilde{p} \leq \tilde{p}_{2} \\
   0  \, , &  \, \, \, {\rm elsewhere}
\end{array}
\right.  \, ,
\label{eq4c-1} \\
&&\hspace{-5mm}
 A_{p} = \frac{(a_{p}-1)}{\tilde{p}_{1}^{1-a_{p}} - \tilde{p}_{2}^{1-a_{p}}}  \, ,
\label{eq4c-2}
\end{eqnarray}
where Eq.~(\ref{eq4c-1}) is normalized by $\int_{0}^{\infty} d \tilde{p} \tilde{p}^{2} \tilde{p}_{e}(p)=1$.  It should be noted that the constant $a_{p}$ is related to the constant $\alpha$ of Colafrancesco et al.\cite{cola03, cola09} by $a_{p}=\alpha+2$, where we choose the reported values\cite{cola03} $\alpha$=2.5 and $\tilde{p}_{2}$=1000, and study the $\tilde{p}_{1}$-dependence of the spectral distortion functions. \\
\noindent
(ii) the $E$-power distribution:
\begin{eqnarray}
&&\hspace{-10mm}
\tilde{p}_{e}(E) = A_{\gamma} \, \gamma^{-a_{\gamma}} \, , \, \, \, 0 \leq \tilde{p} < \infty  \, ,
\label{eq4c-3} \\
&&\hspace{-5mm}
 A_{\gamma} = \frac{4}{\sqrt{\pi}} \frac{\Gamma \left(\displaystyle{\frac{a_{\gamma}}{2}} \right)}{\Gamma \left(\displaystyle{\frac{a_{\gamma}-3}{2}} \right)}  \, ,
\label{eq4c-4}
\end{eqnarray}
where $a_{\gamma}>3$, and Eq.~(\ref{eq4c-3}) is also normalized by $\int_{0}^{\infty} d \tilde{p} \tilde{p}^{2} \tilde{p}_{e}(p)=1$.  In the present paper, we fix the parameter value $a_{\gamma}$=4.5 for an illustrative purpose.

Using Eqs.~(\ref{eq4c-1}) and (\ref{eq4c-3}), one can calculate the normalization of the pressure distribution $P_{1,K}(s)$.  One obtains
\begin{eqnarray}
&&\hspace{-15mm}
\int_{-\infty}^{\infty}ds P_{1,K}(s) = \left\{
\begin{array}{ll}
 \frac{1}{3} a_{p}  \, , &  \, p{\rm-power \, \, distribution} \\
   1  \, , &  \, E{\rm-power \, \, distribution}
\end{array}
\right.  \, ,
\label{eq4c-5} 
\end{eqnarray}
where Eq.~(\ref{eq3b-1}) was satisfied for the $E$-power distribution case.

  First, in Figs.~4a and 4b, we plot $P_{1}(s)$ and $P_{1,K}(s)$, respectively.  The solid curve, dash-dotted curve, and dashed curve correspond to the $p$-power distributions for $\tilde{p}_{1}$ = 0.1, 0.3, and 1.0, respectively.  The dotted curve is the $E$-power distribution.  In Fig.~4b, the dotted curve is multiplied by a factor 10 in order to be visible in the same figure.  It is seen from Fig.~4a that $P_{1}(s)$ for $\tilde{p}_{1}$=0.1 has a sharp peak at $s$=0 similar to the case of $k_{B}T_{e}$=5keV in Fig.~1a, and the peak becomes flatter and wider as $\tilde{p}_{1}$ grows.  Moreover, the peak position is shifting toward the positive direction.  As seen from Fig.~4b, the structure of $P_{1,K}(s)$ is similar to the thermal distribution case, however, the parameter dependence is quite large.

  Similarly, in Figs.~5a and 5b, we plot $\Delta I_{SZ}(x)$ and $\Delta I_{kSZ}(x)$ for $\tau$=0.01, respectively.  The solid curve, dash-dotted curve, and dashed curve correspond to the $p$-power distributions for $\tilde{p}_{1}$ = 0.1, 0.3, and 1.0, respectively.  The dotted curve is the $E$-power distribution.  The curves have long tails even for a small $\tau$ values.  More importantly, it is seen from Fig.~5b that $\Delta I_{kSZ}$ shows a complex structure (two peaks) in the case of the $E$-power distribution.  It is found that the similar structure appears for the $p$-power distributions for $\tilde{p}_{1}>0.8$.  This suggests that data for the kinematical SZ effect might give new limits to the parameter values for the nonthermal distributions provided that the separation of the main SZ effect has been carefully done, for example, by taking into account for the non-isothermal profiles of the CG\cite{hans04b}.

\bigskip

\section{Concluding Remarks}

  Starting from the rate equations for the photon distribution function which was derived in the NK paper, we derived the formal solutions for the SZ effect in three different representation forms: the multiple scattering representation, operator representation, and Fourier transform representation.  In particular, we showed that these representation forms were identical.  By expanding the formal solution in the operator representation in powers of both the derivative operator $D$ and electron velocity $\beta$, we obtained the same formal solution derived in the Fokker-Planck expansion approximation.

   We also extended the present formalism to the kinematical SZ effect.  We obtained the formal solutions in the operator representation.  Analytical properties for the photon frequency redistribution functions were studied.  We found that the redistribution function which corresponds to the kinematical SZ effect can be interpreted as the pressure distribution of electrons.

  We solved the rate equation numerically, and obtained the exact numerical solutions for the thermal SZ effect and kinematical SZ effect.  The exact numerical solutions include the full-order terms in powers of $\tau$, where $\tau$ is the optical depth.  We compared the present solutions with other calculations such as first-order calculation in $\tau$.  As far as the clusters of galaxies are concerned (, where $\tau \ll 1$ is satisfied), the existing calculations are accurate enough to analyze the observational data.  On the other hand, the exact calculation which includes the full-order terms in $\tau$ is necessary for $\tau > 1$.

  Finally, we calculated the spectral intensity functions for the nonthermal electron distributions.  The parameter dependences were studied for the $p$-power distribution.  It was suggested that observation of the kinematical SZ effect might provide new limits to the parameter values for the nonthermal distributions.

\begin{acknowledgments}
This work is financially supported in part by the Grant-in-Aid of Japanese Ministry of Education, Culture, Sports, Science, and Technology under the contract \#21540277.  We would like to thank our referee for valuable suggestions.
\end{acknowledgments}



\bibliography{apssamp}

\end{document}